\documentclass{llncs}
\usepackage{makeidx}  
\usepackage{setspace}
\usepackage{amssymb}
\usepackage{amsmath}
\usepackage{amsfonts}
\usepackage{amssymb}
\usepackage{epsfig}
\usepackage{graphicx}

\usepackage{enumerate}
\usepackage{subfigure}
\usepackage[nosepfour,warning,np,debug,autolanguage]{numprint}
\usepackage[ruled,vlined]{algorithm2e}
\begin{document}
\frontmatter          
\pagestyle{headings}  

\title{Computationally Efficient Technique for Nonlinear Poisson-Boltzmann equation}
%
\titlerunning{Poisson Boltzmann Equation}  
%
\author{Sanjay Kumar Khattri}
%
\authorrunning{S. K. Khattri et al.}   
%
%
\institute{Department of Mathematics, University of Bergen, Norway.\\
\email{sanjay@mi.uib.no}\\
\texttt{{http://www.mi.uib.no/$\sim$sanjay}}
}

\maketitle              
\begin{abstract}
Discretization of non-linear Poisson-Boltzmann Equation equations results in a system of non-linear equations with symmetric Jacobian. The Newton algorithm is the most useful tool for solving non-linear equations. It consists of solving a series of linear system of equations (Jacobian system). In this article, we adaptively define the tolerance of the Jacobian systems. Numerical experiment shows that compared to the traditional method our approach can save a substantial amount of computational work. The presented algorithm can be easily incorporated in existing simulators.
\end{abstract}
%
\section{Introduction}
\label{sec:intro}
Lets consider the following non-linear elliptic problem
\begin{equation}
-\,\text{div}\,(\epsilon\,\text{grad}\,p) + f(p,x,y) = b(x,y) \quad \text{in} \quad \Omega\quad\text{and}\quad
p(x,y) = p^D             \quad  \text{on} \quad  \partial{\Omega_D}\enspace.
\label{elliptic1}
\end{equation}
The above problem is the Poisson-Boltzmann equation arising in molecular bio-physics. See the References \cite{burak,fogolari,holst1,holst_44,holst_55,holst_33}. Here, $\Omega$ is a polyhedral domain in $\mathbb{R}^{2}$, the source function $b$ is assumed to be in $L^2(\Omega)$ and the medium property ${\epsilon}$ is uniformly positive. 

A Finite Volume discretization of the nonlinear elliptic equation results in a system of non-linear equations
\begin{equation}
\mathbf{F(p)} := \mathbf{A}_1 \,\mathbf{p}_h + \mathbf{A}_2(\mathbf{p}_h)-\mathbf{b}_h=0\enspace.
\label{eq:discrete}
\end{equation}
Here, $\mathbf{F}=\left[F_1(\mathbf{p}), F_2(\mathbf{p}), \cdots, F_n(\mathbf{p})\right]^T$, $\mathbf{A}_1$ is the discrete representation of the symmetric continuous operator $-\text{div}\,(\epsilon\,\text{grad})$ and $\mathbf{A}_2$ is the discrete representation of the non-linear operator $f(p,x,y)$. 

A Newton-Krylov method for solving the non-linear equation \eqref{eq:discrete} is given by the Algorithm \ref{algorithm_2}.
\begin{algorithm1}
\label{algorithm_2}
\SetLine 
{
Mesh the domain;\\
Form the non-linear system$\colon$ $\mathbf{F}(\mathbf{p})$;\\
Set the iteration counter$\colon$ $k$ = 0 ;\\
\While{${k} \le  {\max}_{iter}\: or \:\Vert{\Delta{\mathbf{p}}}\Vert_{L_2} \le tol \: or \:
\Vert{\mathbf{F(\mathbf{p})}}\Vert_{L_2} \le tol
$}{
Solve the discrete system $\colon$ $\boldsymbol{J}(\mathbf{p}_{k}) \,\Delta{\mathbf{p}} = -\mathbf{F(\mathbf{p_k})}$ with a fixed tolerance;\\
$\mathbf{p_{k+1}} = \mathbf{p_{k}}+\Delta{\mathbf{p}}$; \\
$k^{++}$;
}
}
\caption{Newton-Krylov Algorithm}
\end{algorithm1}
In the Quasi-Newton method (see Algorithm \ref{algorithm_1}), we are solving the Jacobian equation ($\boldsymbol{J}(\mathbf{p}_{k}) \,\Delta{\mathbf{p}} = -\mathbf{F(\mathbf{p_k})}$) approximately. We are solving the system $\boldsymbol{J}(\mathbf{p}_k)\,\Delta\mathbf{p}_k=-\mathbf{F}(\mathbf{p}_k)+\mathbf{r}_k$ with $\Vert\mathbf{r}_k\Vert$ is chosen adaptively. The quasi-Newton iteration is given by the Algorithm \ref{algorithm_1}. 
\begin{algorithm1}
\label{algorithm_1}
\SetLine 
{
Mesh the domain;\\
Form the non-linear system$\colon$ $\mathbf{F}(\mathbf{p})$;\\
Set the iteration counter$\colon$ $k$ = 0;\\
\While{$ {k} \le  {\max}_{iter}\: or \:\Vert{\Delta{\mathbf{p}}}\Vert_{L_2} \le tol \: or \:
\Vert{\mathbf{F(\mathbf{p})}}\Vert_{L_2} \le tol$}{
Solve the discrete system $\colon$ $\boldsymbol{J}(\mathbf{p}_{k}) \,\Delta{\mathbf{p}} = -\mathbf{F(\mathbf{p_k})}$ with a tolerance $\numprint{1.0\times{10}^{-({k+1})}}$;\\
$\mathbf{p_{k+1}} = \mathbf{p_{k}}+\Delta{\mathbf{p}}$; \\
$k^{++}$;
}
}
\caption{Quasi-Newton-Krylov Algorithm}
\end{algorithm1}
In the Algorithms \ref{algorithm_2} and \ref{algorithm_1}, $\Vert\cdot\Vert_{{L}_2}$ denotes the discrete $L_2$ norm and max$_{{iter}}$ is the maximum allowed Newton's iterations. It is interesting to note the stopping criteria in the Algorithm \ref{algorithm_1}. We are using three stopping criterion in the Algorithms. Apart from the maximum allowed iterations, $L_2$ norm of residual vector ($\Vert{ \mathbf{F(\mathbf{p})}}\Vert_{{L}_2}$) and also $L_2$ norm of difference in scalar potential vector ($\Vert{\Delta{\mathbf{p}}}\Vert_{{L}_2}$) are being used as stopping criterion for the Algorithms. Generally in the literature, maximum allowed iterations and the residual vector are used as stopping criteria \cite[and references therein]{holst1,holst_44,holst_55}. If the Jacobian is singular than the residual vector alone cannot provide a robust stopping criteria. 
\section{Numerical Experiment}
\label{sec:numerical_work}
\begin{figure}
\begin{minipage}[b]{0.5\linewidth} 
\centering
\includegraphics[scale=0.5]{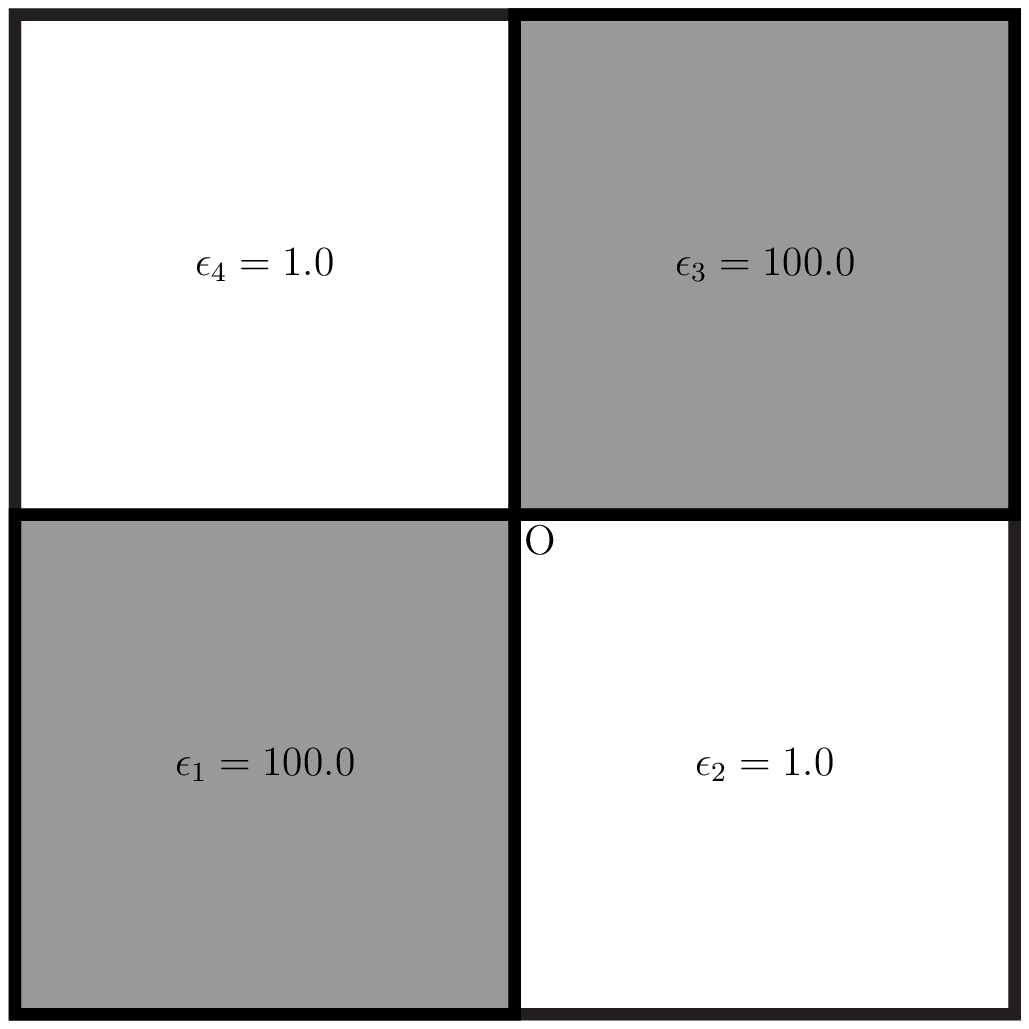}
\caption{Distribution of medium property $\epsilon$ in the domain $\Omega=[-1,1]\times[-1,1]$.}
    \label{fig:domain_1}
\end{minipage}
\hspace{0.5cm} 
\begin{minipage}[b]{0.5\linewidth}
\centering
\includegraphics[scale=0.5]{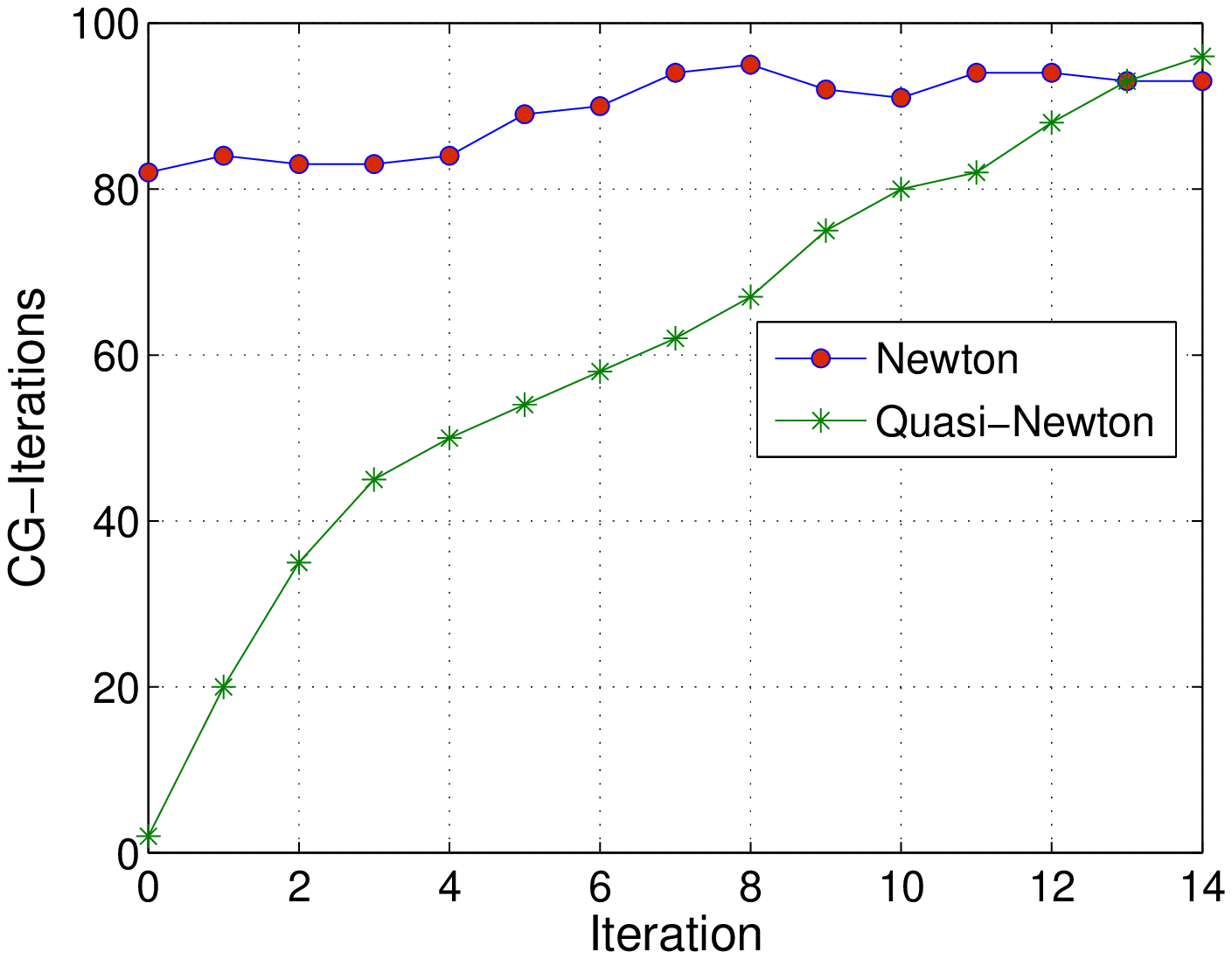} 
\caption{Computational work required by the Quasi-Newton and Newton methods.}
 \label{fig:newton_cg_00}
\end{minipage}
\end{figure}

\begin{figure}
\begin{minipage}[b]{0.5\linewidth} 
\centering
\includegraphics[scale=0.40]{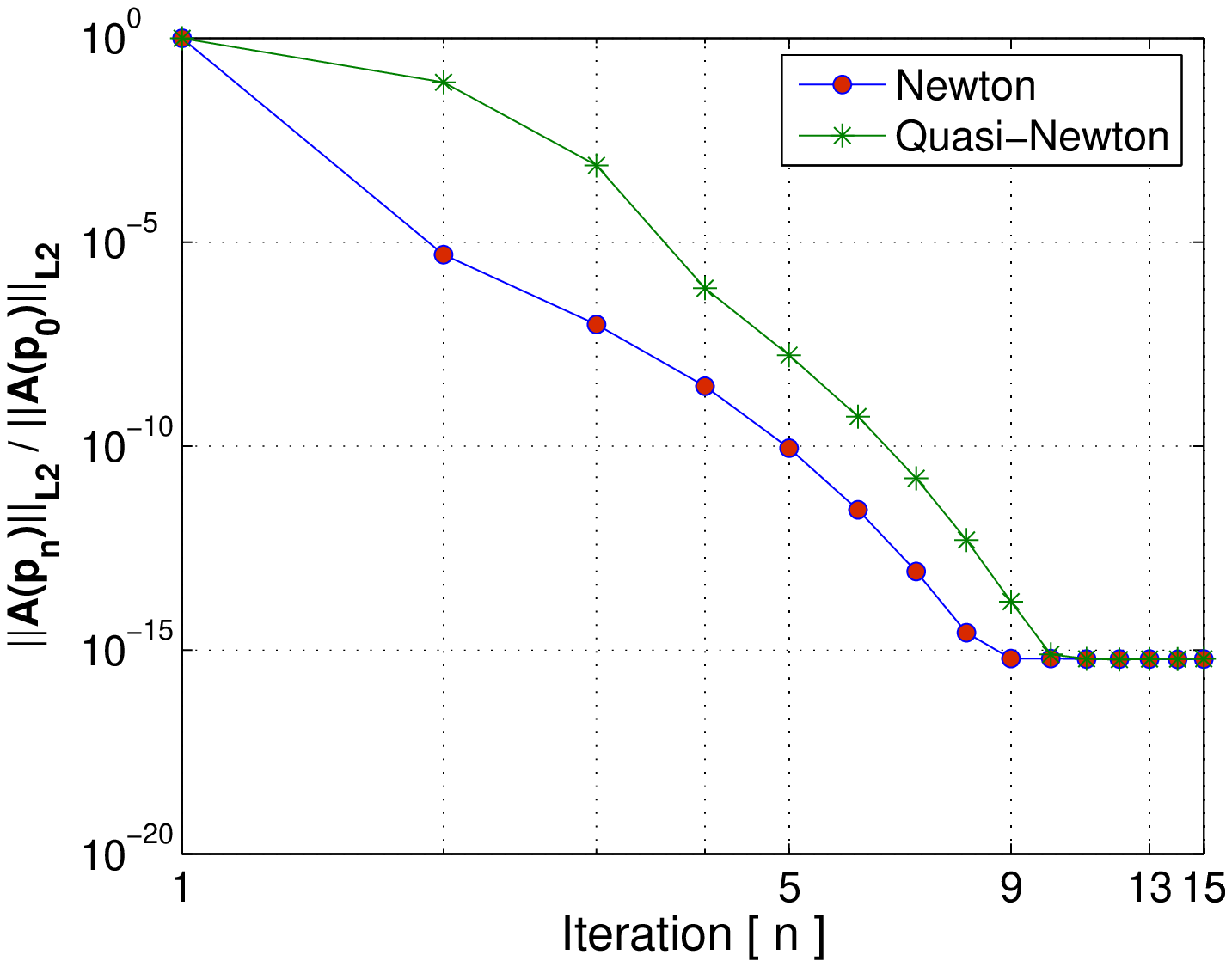}
\caption{Convergence of the $L_2$ norm of residual vector $\mathbf{A}(\mathbf{p})$.}
    \label{fig:ap_00}
\end{minipage}
\hspace{0.5cm} 
\begin{minipage}[b]{0.5\linewidth}
\centering
    \includegraphics[scale=0.40]{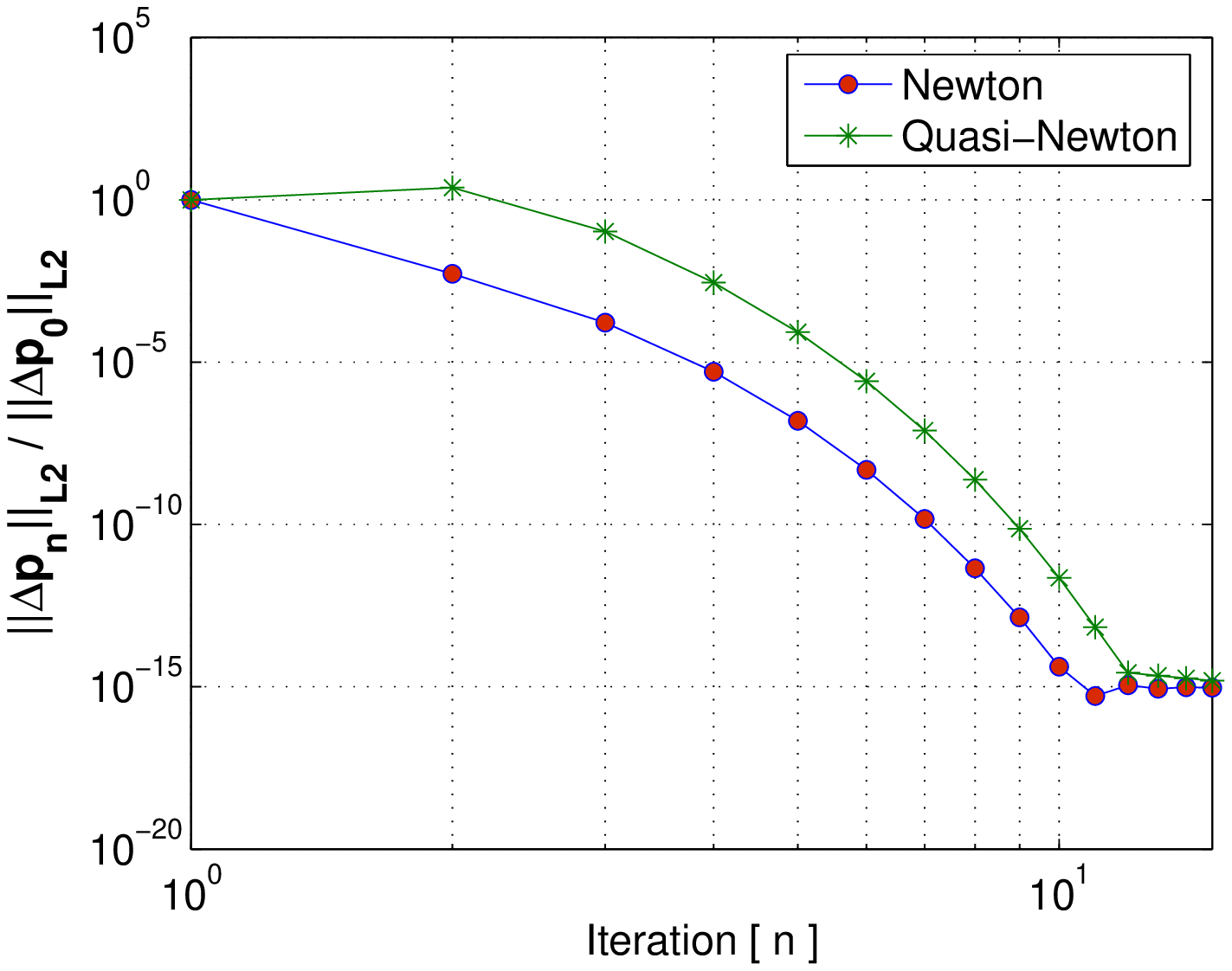}
\caption{Convergence of the $L_2$ norm of difference vector $\Delta\mathbf{p}$.}
    \label{fig:delp_00}
\end{minipage}
\end{figure}

Let us solve \eqref{elliptic_300} in the domain $\Omega=[-1,1]\times[-1,1]$ with $k=1.0$ \cite{burak,fogolari,jacob1,holst1}. $\Omega$ is divided into four equal sub-domains (see Figure \ref{fig:domain_1}) based on $\epsilon$. 
\begin{equation}
-\,\nabla\cdot({\epsilon\,\nabla{p}})+k\,\sinh(p) = f \quad \text{in} \quad \Omega\quad\text{and}\quad p(x,y) = x^3+y^3             \qquad  \text{on} \quad  \partial{\Omega_D}\enspace.\label{elliptic_300}
\end{equation}
For solving the linear systems, we are using ILU-preconditioned the Conjugate-Gradient (CG) method. For the Newton algorithm the tolerance of the CG method is $\numprint{1.0\times{10}^{-15}}$. For the quasi-Newton method the tolerance of the CG method varies with the iterations $k$ of the Algorithm \ref{algorithm_1} as follows$\colon$ $\numprint{1.0\times{10}^{-(k+1)}}$, $k=0,2,\ldots,14$. Figures \ref{fig:ap_00}, \ref{fig:delp_00} and \ref{fig:newton_cg_00} reports the outcome of our numerical work. The Figures \ref{fig:ap_00} and \ref{fig:delp_00} compares convergence of the quasi-Newton and Newton methods. The Figure \ref{fig:newton_cg_00} reports computational complexity of the quasi-Newton and the Newton methods. It can be notice, even if initial iterations of the Newton-Krylov algorithm are solved approximately, the convergence rate of the algorithm remains unaffected. The Figure \ref{fig:newton_cg_00} shows that such an approximation saves a substantial amount of computational effort.   
\section{Conclusions}
\label{sec:conclusions}
Quasi-Newton method for solving non-linear system of equation with symmetric Jacobian matrix is presented. Numerical work shows that the presented technique is computationally efficient compared to the traditional Newton-Krylov method. An efficient solution technique for Poisson-Boltzmann equation is of interest to the researchers in computational chemistry, bio-physics and molecular dynamics. The presented algorithm can be easily implemented in existing simulators.
%
%

%
%
\end{document}